\documentclass{article}

\usepackage{arxiv}

\usepackage[utf8]{inputenc} 
\usepackage[T1]{fontenc}    
\usepackage{hyperref}       
\usepackage{url}            
\usepackage{booktabs}       
\usepackage{amsfonts}       
\usepackage{nicefrac}       
\usepackage{microtype}      
\usepackage{lipsum}
\usepackage{graphicx}
\graphicspath{ {./images/} }
\usepackage{marginnote}
\usepackage{amsmath, amsthm, amscd, amsfonts, booktabs, color, float, geometry, graphicx, tabularx, url, xcolor} 
\usepackage{subcaption}
\usepackage{todonotes}

\title{Aspects of complexity in automotive software systems and their relation to maintainability effort. A case study}

\author{
 Bengt Haraldsson \\
  Chalmers University of Technology and \\ University of Gothenburg\\
  Gothenburg\\
  Scania CV AB \\
  Södertälje \\
  \texttt{bengthar@chalmers.se} \\
   \And
 Miroslaw Staron \\
  Chalmers University of Technology and \\ University of Gothenburg\\
  Gothenburg\\
  \texttt{miroslaw.staron@gu.se} \\
   \\
}

\begin{document}
\maketitle
\begin{abstract}
\textbf{Context}: Large embedded systems in vehicles tend to grow in size and complexity, which causes challenges when maintaining these systems. 
\textbf{Objective}: We explore how developers perceive the relation between maintainability effort and various sources of complexity. 
\textbf{Methods}: We conduct a case study at Scania AB, a heavy vehicle OEM. The units of analysis are two large software systems and their development teams/organizations. 
\textbf{Results}: Our results show that maintainability effort is driven by system internal complexity in the form of variant management and complex hardware control tasks. The maintainability is also influenced by emergent complexity caused by the system's longevity and constant growth.  Besides these system-internal complexities, maintainability effort is also influenced by external complexities, such as organizational coordination and business needs. During the study, developer trade-off strategies for minimizing maintainability effort emerged.
\textbf{Conclusions}: Complexity is a good proxy of maintainability effort, and allows developers to create strategies for managing the maintainability effort. Adequate complexity metrics include both external aspects---e.g., coordination complexity---and internal ones---e.g., McCabe Cyclomatic Complexity.  
\end{abstract}


\section{Introduction}
\label{sec:introduction}
\noindent
Embedded software in the automotive domain becomes increasingly complex because of the constant growth of functionality; heavy vehicles like trucks are examples of such products \cite{staron2021automotive}. The growing complexity leads to increased effort to maintain the software. Automotive OEMs address the problem of increased complexity by adopting modern software development paradigms like SAFe and technologies like Continuous Integration/Continuous Deployment \cite{weller2023developing}. 

However, ultimately, the increased effort to maintain software falls on the software developers, architects, and testers who must add new functions to the existing code base while maintaining its overall quality, reliability, and safety to keep the software maintainable. According to ISO/IEC 25010:2023 \cite{ISO}, maintainability can be approximated using such metrics as complexity and readability. The maintainability of software systems has been a cornerstone of software engineering research due to its profound impact on long-term system evolution and costs. Despite the availability of various metrics to evaluate maintainability, discrepancies often arise between calculated metrics and developers' perceptions of maintainability. For example, many studies use McCabe complexity as a proxy for maintainability, but complex software is only unmaintainable if it is not balanced by the competence of the team that develops it \cite{anda2007assessing}. Additionally, maintainability is influenced by system architecture, team structure, and the dynamic interplay between these factors.

Understanding these discrepancies is critical for developing effective tools and practices that align technical evaluations with software engineering practice. Therefore, in this paper, we conduct a case study (according to the guidelines of Runeson and Höst \cite{runeson2009guidelines}) at one of the Swedish vehicle manufacturers, Scania CV AB, which develops complex embedded software for large organizations. The software is subject to rigorous standardization and certification, which requires high maturity in its development processes. At the same time, the software product has grown and become increasingly complex, causing challenges for its maintainability and timely delivery.

\begin{enumerate}
    \item \textbf{RQ1:} Which aspects of complexity do developers highlight regarding maintainability effort?
    \item \textbf{RQ2:} How well do code metrics and variant complexity align with a developer-experienced maintainability effort?
    \item \textbf{RQ3:} How do team structure and interactions align with developer-experienced maintainability effort?
\end{enumerate}

We combine a metric-based analysis with qualitative developer interviews to provide an understanding of maintainability that bridges the gap between technical and human-centered perspectives. Our results show that developers make trade-off decisions that allows for minimizing the total effort of maintainability, while simultaneously managing the risk of change. This means that developers will sometimes go against commonly held beliefs of best-practice in the source-code to minimize risk or lower the effort of maintaining software.

The remainder of the paper is structured as follows. Section \ref{sec:theoretical_framework} explains the main concepts studied and the relationships between them. Section \ref{sec:related_work} introduces the most recent studies in this area. Section \ref{sec:methodology} describes the details of the research design of this study. Section \ref{sec:results} presents the results, and Section \ref{sec:discussion} discusses them in the light of the existing studies. Finally, Section \ref{sec:conclusions} presents the conclusions drawn from this study. 

\section{Theoretical Framework}
\label{sec:theoretical_framework}
\begin{figure}
    \centering
    \includegraphics[width=0.5\linewidth]{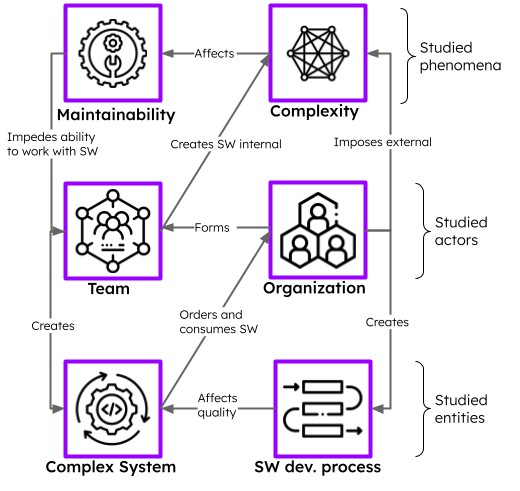}
    \caption{Theoretical Framework}
    \label{fig:theoretical_FW}
\end{figure}

In the context of this study, the theoretical framework applied when constructing the research is depicted in Figure \ref{fig:theoretical_FW}. We are interested in the high-level concepts of maintainability and complexity. Maintainability is a quality property that is affected by complexity, among other factors. Although it is an intangible property that is hard to measure quantitatively, it is recognizable by most developers. At the same time, the complexity of the software is more tangible but still a multifaceted concept. It is affected by both the complexity of the problem addressed in the software and the team that develops it. A complex problem often leads to complex software. 

Complexity is viewed as comprised of several interrelated aspects of a whole, including code complexity, product complexity, and organizational complexity. Code complexity can be measured by different complexity metrics, product complexity is made up of intricacies related to the problem domain---such as e.g., variant management. These are interrelated in the sense that they influence each other. For example, product complexity can drive design choices that increases code complexity. Code complexity and product complexity together form a complex system.

The organization itself often drives the team and the complexity of the problem. The organization decides how the team is structured and which competence it has. The software development teams and organizations are more tangible than maintainability and complexity properties, but they are still more elusive as the entities like software source code, variants, or software development processes followed.  

Therefore, this study also focuses on the tangible elements of software development organizations—software products (we focus on the source code and variants) and software development processes---de facto ways of working in the organization.  

The conceptual model depicted in Figure \ref{fig:theoretical_FW}, shows the interactions between studied phenomena, actors, and entities. This model is a sub-set of the total set of phenomena, actors, and entities that could have been studied. The sub-set was chosen as the one most applicable for this study. We focus on the complexity metrics as the proxy of maintainability because we can reason about increasing/decreasing complexity \cite{Antinyan_revealing}. As previous studies indicated, complexity and size often correlate with many other properties \cite{Mamun}, which allows us to use it as a proxy for many of them.

\section{Related Work}\label{sec:related_work}
Maintainability is a multifaceted concept that has been the focus of extensive research in software engineering. Coleman et al.~\cite{Coleman} introduced foundational metrics for evaluating maintainability, emphasizing the importance of code structure and readability. Subsequent work by Antinyan et al.~\cite{Antinyan_maintainance_time, Antinyan_hypnotized} expanded on this by exploring the impact of complexity triggers and practitioners' perceptions on maintenance activities. Heitlager et al.~\cite{Heitlager} proposed practical models for measuring maintainability, highlighting the challenges of applying theoretical metrics in real-world scenarios.

Studies examining the relationship between maintainability metrics and developer perceptions have revealed areas where code metrics do not capture the whole developer experience. Holzmann~\cite{Holzmann} and Buse and Weimer~\cite{Buse} emphasized that metrics often fail to capture nuances such as code readability and cognitive load, which are critical to developers. Similarly, Mamun et al.~\cite{Mamun} highlighted the limitations of automated measurements in capturing the socio-technical aspects of maintainability. Developer perceptions have been used to assess system defects, e.g., Santos et al.~\cite{Santos} used developer perception to evaluate an AI-tool generated defect predictability assessment, finding that the developers focused on code complexity over other metrics.

The interplay between team structure and system architecture has also been explored, with Lehman~\cite{Lehman} and Bosch and Bosch-Sijtsema~\cite{Bosch} providing insights into how organizational dynamics influence software evolution. Concas et al.~\cite{Concas} and Mubarak et al.~\cite{Mubarak} further examined social network metrics in software teams, demonstrating the impact of collaboration patterns on code quality and maintainability. 

Conejero et al.~\cite{Conejero} examined the relation between maintainability and technical debt in the requirement level, relating these to modularity. They found that a larger amount of what they call \textit{modularity anomalies}, the greater the maintainability effort. Related to this, Gordieiev et al.~\cite{Gordieiev} studied the long-term evolution of a system. They found that the relation between the software development process and requirement handling has effects on the amount of defects in a system.

The study of cyclomatic complexity and other metrics as measures of system complexity, and its impact on maintainability has been studied to a great extent, e.g., complexity~\cite{Gill} and information flow~\cite{Henry}. Developer perceptions have been used to assess system defects, e.g., Santos et al.~\cite{Santos} used developer perception to evaluate an AI-tool generated defect predictability assessment, finding that the developers focused on code complexity over other metrics.

A number of systematic literature reviews have been produced on the topic of maintainability. Mehwish et al.~\cite{Mehwish} investigated the predictive power of code metrics, finding little evidence of their effectiveness. Ardito et al.~\cite{Ardito} examined tooling for code metric analysis, identifying an optimal set of tools, but also identifying missing coverage for some cases---such as for some programming languages. Abílio et al. investigated feature- and aspect-oriented programming and constructed a lists of metrics for these technologies. Elmidaoui et al.~\cite{Elmidaoui} collected and evaluated research on software maintainability prediction comparing the accuracy of different models, and concluding that the models are still limited. Malhotra and Anuradha \cite{Malhotra} investigated trends in software maintainability prediction, finding that machine learning approaches have increased, but concluding that hybrid techniques are preferable and that more empirical studies are needed in the field. Finally, in a later review Malhotra and Lata \cite{Malhotra2} found that machine learning approaches to maintainability predictions outperformed statistical models, but that hybrid techniques were still rarely used.

While these studies offer valuable insights, they often treat maintainability metrics, developer perceptions, and organizational factors as separate domains. This study builds on this foundation by investigating the alignment and interactions between these elements with the aim to provide a more holistic understanding of software maintainability effort.

\section{Methodology}
\label{sec:methodology}
Our case study adopts a mixed methods approach, integrating quantitative analysis of software metrics and system structure with qualitative insights from developers. The methodology is structured around a case study characterized as a \textit{positivist explanatory} \cite{Runeson} seeking to explain the link between code metrics and developer-experienced maintainability effort---and how these relate to organization and architecture. Thus, the \textit{phenomena under study} are maintainability and complexity.

Following Antinyan et al.~\cite{Antinyan_functions}, quantitative data on code complexity was gathered using \textit{cyclomatic complexity}, \textit{structural fan-in}, and \textit{structural fan-out} metrics. Lines of code without blank spaces and comments were also calculated to provide general information about the systems under study. Other metrics were excluded due to their correlation with cyclomatic complexity \cite{Antinyan_functions}. 

Qualitative data was gathered via interviews with five senior developers and one meeting with a technical manager to gather overall information such as team organization charts and release planning process description. Information about the developers can be found in Table \ref{tab:developers}.

\begin{table}[htbp]
    \centering
    \begin{tabular}{|c|c|c|c|}
    \hline
    \textbf{Developer} & \textbf{Current system} & \textbf{Total} & \textbf{System}\\
    \hline
    1A & 7 & 23 & A\\
    \hline
    2A & 16 & 20+ & A\\
    \hline
    3A & 12 & 12 & A\\
    \hline
    1B & 6 & 6 & B\\
    \hline
    2B & 7 & 18 & B\\
    \hline
    \end{tabular}
    \caption{Developer years of experience.}
    \label{tab:developers}
\end{table}

The \textit{methodological triangulation} of combining qualitative and quantitative data was augmented using \textit{data (source) triangulation} combining an investigation of two real-time embedded systems developed in different parts of the organization. This means that there are two \textit{units of analysis} in the case, i.e., two system-developers-units within the same \textit{context}.

Integrating the quantitative metrics and qualitative interviews enabled an examination of maintainability from both technical and human-centered perspectives. Triangulation of quantitative and qualitative findings ensured robustness and validity in addressing the research questions.

\subsection{Case Description}
The case study is set within Scania CV AB, an Original Equipment Manufacturer (OEM) specializing in the heavy vehicle sector of the automotive industry. Historically, Scania has utilized a combination of in-house and third-party software, with a stronger emphasis on in-house development compared to many other OEMs. In recent years, as is the general trend within the automotive industry, the organization has experienced significant growth in both the complexity and size of its vehicle software systems. This growth has necessitated a substantial increase in the developer workforce, which in turn has strained traditional interaction patterns—--e.g., social networks and relationships among senior developers. 

Additionally, the organization has expanded its operations to include software delivery to other brands within its corporate group, referred to in this text as \textit{external delivery}. The inclusion of other brands and the historical variety of trucks delivered resulted in a significant number of software variants. The size, domain, and maturity of the organization, make it an exemplary context for a case study.

In this case, the two units of analysis are two real-time, safety-critical embedded control systems. Both systems are developed in-house and operate on the same internally developed platform software, which includes the operating system and middleware for functions such as CAN communication. The ECU hardware and firmware used in these systems are provided by Tier 1 suppliers. These systems have undergone several generational iterations and have been deployed in products for over 20 years. The in-house development and long history made these systems suitable units of analysis for this study. 

\begin{table}[htbp]
    \centering
    \begin{tabular}{|c|c|c|c|}
    \hline
     \textbf{System} & \textbf{\# modules} & \textbf{Approx. \# files} & \textbf{Approx. \# LOC}\\
     \hline
      A & 58 & 2,000 & 500,000\\
      \hline
      B & 121 & 3,000 & 800,000\\
      \hline
    \end{tabular}
    \caption{System descriptions. Numbers are given, excluding platform software.}
    \label{tab:systems}
\end{table}

Each system comprises several combined modules to form different products and use variant handling to handle multiple deployments. Variations in module behavior depend on the overall product configuration, mechatronic hardware installation, or specific hardware components controlled by the software. Variant handling can be achieved within the source code or by creating separate module variants—largely derived from the same code base with minor adjustments. An overview of the systems’ characteristics is provided in Table \ref{tab:systems}.

The developers included in this study were primarily selected based on their seniority. Seniority was a crucial criterion, as it ensured that the developers had the experience and historical knowledge necessary to provide insights into the evolution of maintainability within their respective systems over time. These developers were chosen from 11 development teams that work with these systems, each team ranging between 6-12 people. From these teams the developers were sampled based on recommendations from one senior manager, two managers, and one senior technical advisor.

\subsection{Quantitative Metrics Analysis for Maintainability (RQ2, RQ3)}
Maintainability metrics were calculated according to the specifications in ISO/IEC 25010~\cite{ISO} and models proposed by Heitlager et al.\cite{Heitlager} and Antinyan et al.\cite{Antinyan_maintainance_time}. The metrics were implemented in Python. Cyclomatic complexity was assessed using the Lizard\footnotemark{} Code Complexity calculator\footnotetext{https://github.com/terryyin/lizard}, while fan-in and fan-out metrics were derived by parsing the code with the LLVM\footnotemark{} compiler\footnotetext{https://llvm.org/}. These quantitative metrics were then analyzed alongside qualitative data collected from developer interviews. For the two systems under study, the analysis of System A focused on differences in the hardware it controlled, while the analysis of System B examined the distinctions between internal and external delivery.

The team structure was analyzed using organizational charts and cross-referenced with interview responses to uncover key interaction patterns and their impact on perceived maintainability efforts.

Data on system structure and module variant handling was gathered by analyzing documentation that detailed how various module variants were combined to create different products.

\subsection{Developer Perception Analysis (RQ1, RQ2, RQ3)}
Semi-structured interviews were conducted with developers to gather qualitative data on their experiences and perceptions of maintainability. The prepared interview questions are given in the appendix, Section \ref{sec:appendix_dev_questions}.

The interview data was analyzed using thematic analysis. The analysis investigated how developers articulate the relationship between perceived maintainability effort, the collected code metrics, how interaction patterns influence their maintenance practices, and the relation to module variant complexity.

A systematic process was implemented to ensure data integrity in the analysis of developer interviews. Thematic codes were documented alongside their corresponding excerpts and cross-checked to maintain consistency in interpretation.

\section{Results}
\label{sec:results}
We start our analysis with the quantitative data about the complexity of the software systems A and B. 

\subsection{Code metric analysis}
The cyclomatic complexity, fan-in, and fan-out were calculated for two complete versions of each system---A:1, A:2, B:1, and B:2, referred to as products. The products are, therefore, designated as follows:
\begin{itemize}
    \item \textbf{Product A:1:} Internal delivery, control of hardware H:1
    \item \textbf{Product A:2:} Internal delivery, control of hardware H:2
    \item \textbf{Product B:1:} Internal delivery, control of main hardware H
    \item \textbf{Product B:2:} External delivery, control of main hardware H
\end{itemize}

The internal deliveries were products used by Scania vehicles, whereas the external products were delivered to other brands in the corporate group. 

\subsubsection{Products A:1 and A:2}
\begin{figure}[!htbp]
    \centering
    \begin{subfigure}[t]{0.45\linewidth}
        \centering
        \includegraphics[width=\linewidth]{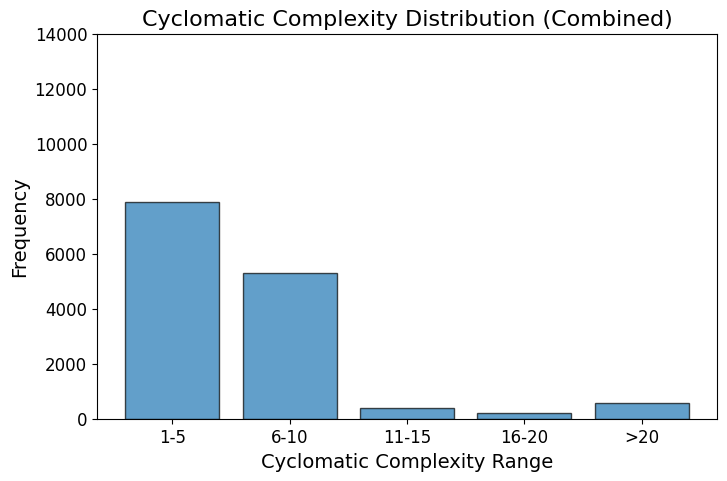}
        \caption{Cyclomatic complexity histogram showing the spread of the metric over functions for product A:1.}
        \label{fig:complex_72_sub}
    \end{subfigure}%
    \hfill
    \begin{subfigure}[t]{0.45\linewidth}
        \centering
        \includegraphics[width=\linewidth]{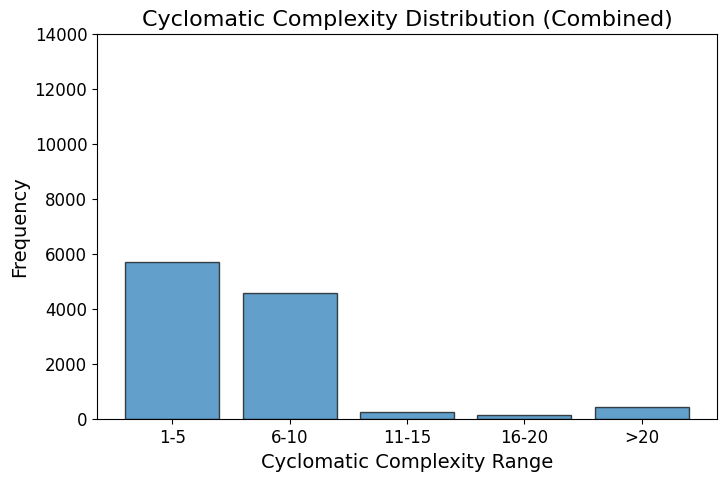}
        \caption{Cyclomatic complexity histogram showing the spread of the metric over functions for product A:2.}
        \label{fig:complex_79_sub}
    \end{subfigure}

    \vspace{1em}
    \begin{subfigure}[t]{0.45\linewidth}
        \centering
        \includegraphics[width=\linewidth]{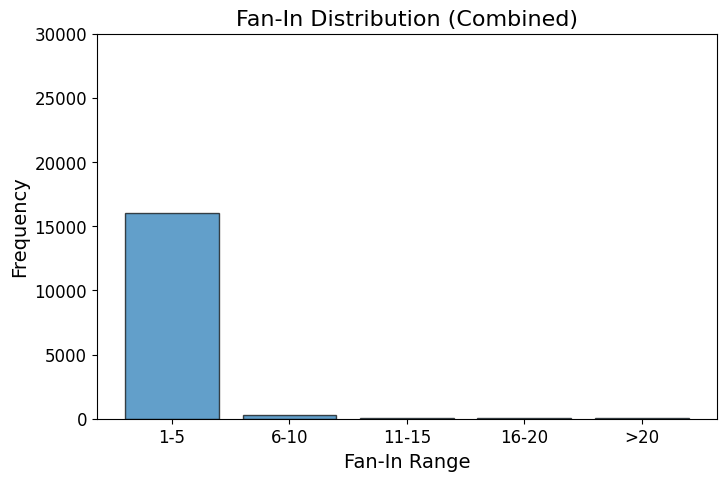}
        \caption{Fan-in histogram for product A:1.}
        \label{fig:fan_in_72_sub}
    \end{subfigure}
    \hfill
    \begin{subfigure}[t]{0.45\linewidth}
        \centering
        \includegraphics[width=\linewidth]{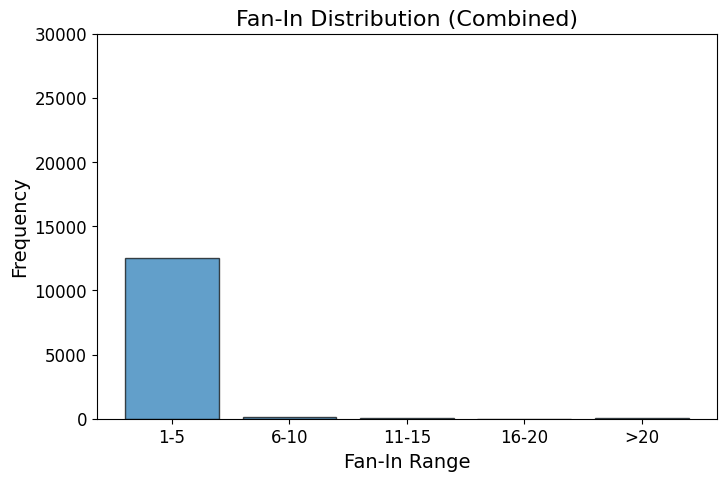}
        \caption{Fan-in histogram for product A:2.}
        \label{fig:fan_in_79_sub}
    \end{subfigure}

    \vspace{1em}
    \begin{subfigure}[t]{0.45\linewidth}
        \centering
        \includegraphics[width=\linewidth]{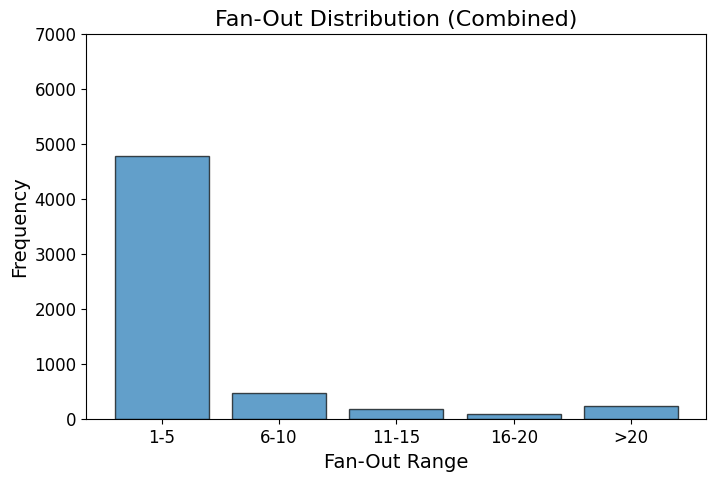}
        \caption{Fan-out histogram for product A:1. Most functions call very few other functions. }
        \label{fig:fan_out_72_sub}
    \end{subfigure}
    \hfill
    \begin{subfigure}[t]{0.45\linewidth}
        \centering
        \includegraphics[width=\linewidth]{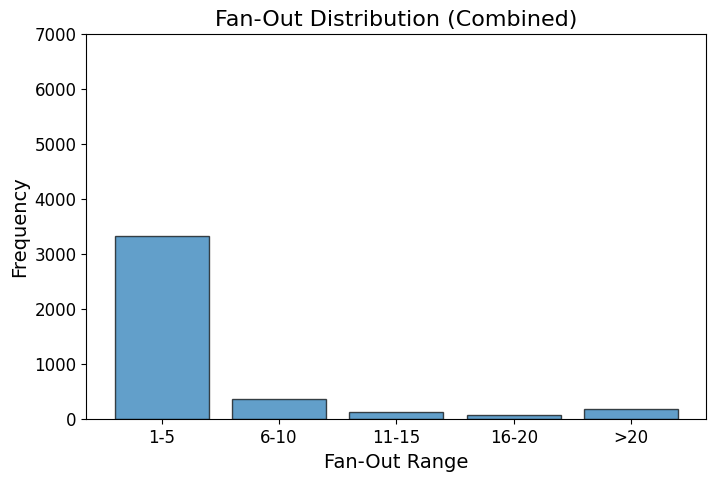}
        \caption{Fan-out histogram for product A:2. The trend is similar to A:1.}
        \label{fig:fan_out_79_sub}
    \end{subfigure}
    
    \caption{Complexity metrics for A:1 and A:2. The diagrams show that the distributions are similar between these two versions, but A:2 is larger in size in terms of a number of functions. }
    \label{fig:comparison_complexity}
\end{figure}

Figure \ref{fig:comparison_complexity} shows the complexity of two systems -- A:1 and A:2. They implement the same functionality but for controlling different hardware versions. Figure \ref{fig:complex_72_sub} and Figure \ref{fig:complex_79_sub} show histograms with similar trends. Most functions have a complexity below 10, and the largest number of functions have a complexity below 5. 

The most complex functions, however, are the most interesting ones since they can indicate challenges with their maintainability. We identified two primary sources of this high complexity: 1) continuous development over a long period and 2) handling of variants inside large, so-called common functions. Both types of complexity relate to design decisions but can cause low maintainability. The complex mechatronic hardware's complex control software must contain complex state machines. The same is true for variants---reusing the same control software for different engines causes multiple variants, which cause higher complexity---\texttt{if} or \texttt{switch} statements. 

A comparison of Products A:1 and A:2 reveals that, although they contain the same number of modules, Product A:1 includes a significantly higher number of functions. While the distributions appear broadly similar, the greater number of functions in Product A:1 suggests that it is inherently larger.

Functions associated with high fan-in were checked manually, with aid from the developers to clarify when needed. Several general categories of functions were identified: 1) functions for getting raw CAN data, 2) helper-functions to calculate physical quantity, 3) helper-functions to check preconditions, 4) utility functions to update values in data structures, 5) system functions to access system information, helper-functions for complicated and re-usable functionality, 6) helper-functions to request actuation of hardware parts, and 7) Simulink generated helper-function to perform a change in set state. 

High scores in fan-out were also checked, and these functions fell into the following identified categories: 1) functions related to real-time database actions calling middleware functions, 2) functions that received information from other systems over CAN, extracted, and transformed the data, 3) functions that process raw data from sensors, 4) periodically executed control functions using feedback control and model-based algorithms, 5) stateful functions that handle errors, translate internal values to outputs, and sent these to other systems, and 6) step functions generated by Simulink called by the scheduler. 

\subsubsection{Product B:1 and B:2}
\begin{figure}[!htbp]
    \centering
    \begin{subfigure}[t]{0.45\linewidth}
        \centering
        \includegraphics[width=\linewidth]{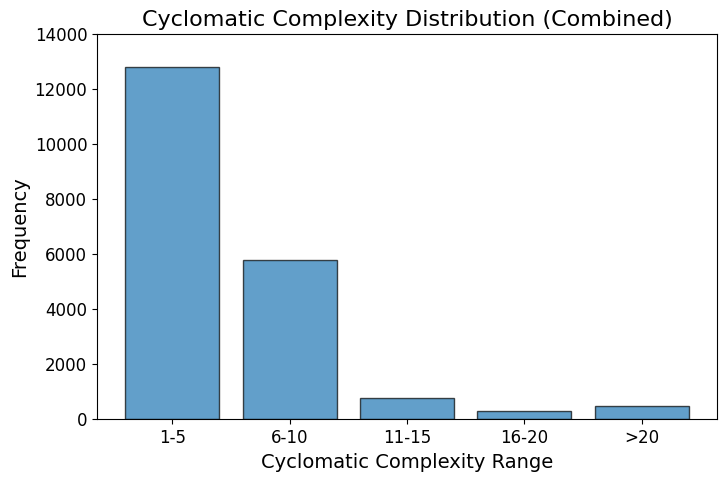}
        \caption{Cyclomatic complexity histogram showing the spread of the metric over functions for product B:1.}
        \label{fig:complex_81_sub}
    \end{subfigure}%
    \hfill
    \begin{subfigure}[t]{0.45\linewidth}
        \centering
        \includegraphics[width=\linewidth]{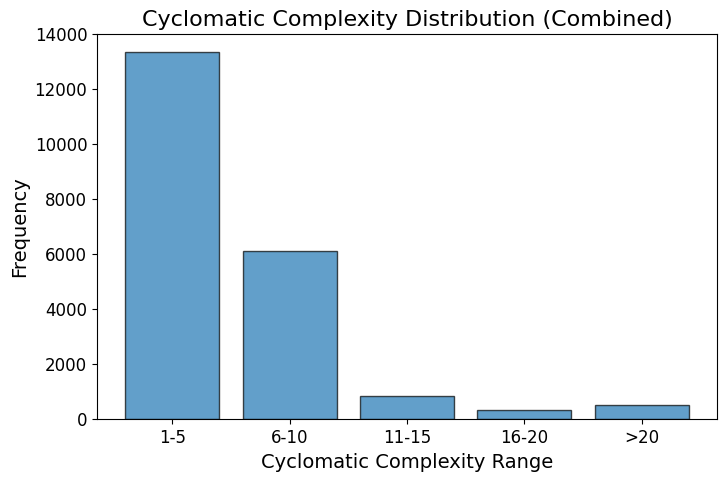}
        \caption{Cyclomatic complexity histogram showing the spread of the metric over functions for product B:2.}
        \label{fig:complex_82_sub}
    \end{subfigure}

    \vspace{1em}
    \begin{subfigure}[t]{0.45\linewidth}
        \centering
        \includegraphics[width=\linewidth]{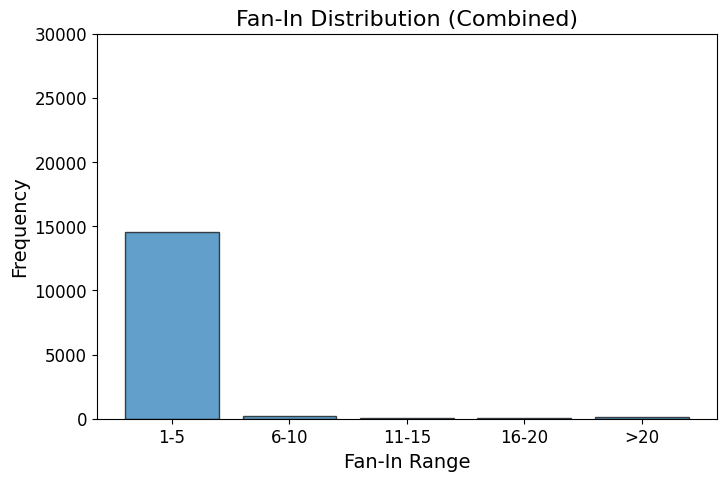}
        \caption{Fan-in histogram for product B:1.}
        \label{fig:fan_in_81_sub}
    \end{subfigure}
    \hfill
    \begin{subfigure}[t]{0.45\linewidth}
        \centering
        \includegraphics[width=\linewidth]{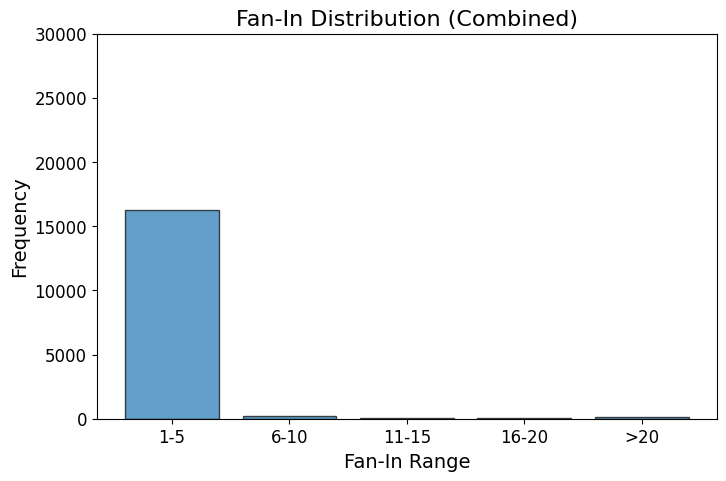}
        \caption{Fan-in histogram for product B:2.}
        \label{fig:fan_in_82_sub}
    \end{subfigure}

    \vspace{1em}
    \begin{subfigure}[t]{0.45\linewidth}
        \centering
        \includegraphics[width=\linewidth]{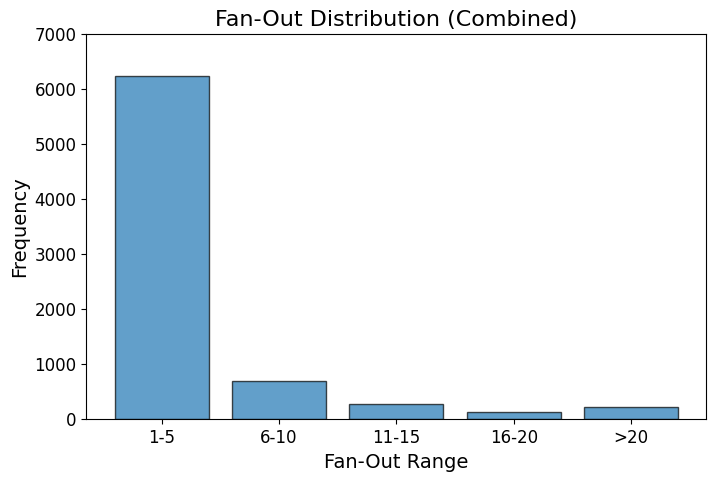}
        \caption{Fan-out histogram for product B:1. Most functions call very few other functions. }
        \label{fig:fan_out_81_sub}
    \end{subfigure}
    \hfill
    \begin{subfigure}[t]{0.45\linewidth}
        \centering
        \includegraphics[width=\linewidth]{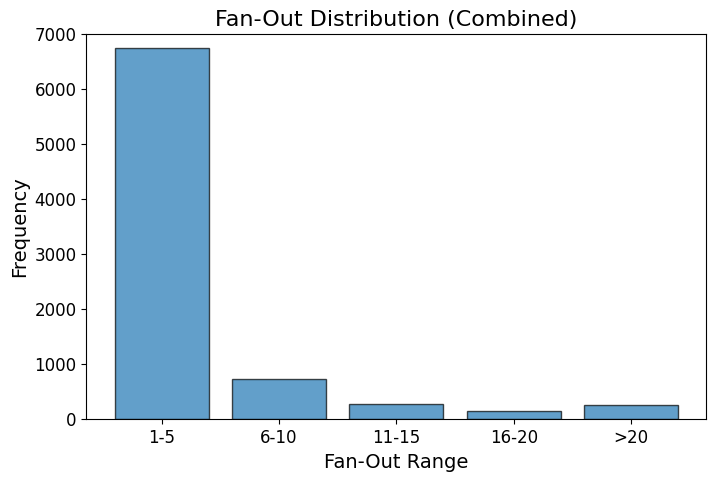}
        \caption{Fan-out histogram for product B:2. The trend is similar to B:1.}
        \label{fig:fan_out_82_sub}
    \end{subfigure}
    
    \caption{Complexity metrics for B:1 and B:2. The diagrams show that the distributions are similar between these two versions, indicating that external deliveries do not influence product complexity to any greater extent.}
    \label{fig:comparison_complexity_second}
\end{figure}

A comparison of Product B:1---see Figure \ref{fig:comparison_complexity_second}---with the previous two products reveals that this system contains a significantly higher number of functions. This observation is consistent with the larger size of Product B:1, as measured by lines of code (LOC) and the number of modules. Notably, the fan-in metric is more consistently distributed between 1 and 5, whereas the fan-out metric exhibits slightly higher values. The number of functions with a complexity exceeding 20 is comparable. This suggests the presence of a core set of functions that inherently require higher complexity.

A comparison of Products B:1 and B:2 reveals that they are highly similar, with no significant differences observed from a product metric perspective. This suggests that, in contrast to Product A, controlling different hardware configurations can contribute to increased internal product complexity and maintenance effort. However, external delivery does not significantly affect the product metrics.

This system's main source of complexity was, in part, the same as for System A 1) and 2), with the addition of  3) state-full functions for managing hardware tasks. 

Functions with high fan-in values were found to fall into the following categories: 1) functions for managing input/output towards hardware, 2) helper functions for various messaging, 3) conversion functions between physical properties, 4) monitoring and state-information functions, and 5) functions for storing data.

Fan-out heavy functions fell into the categories: 1) checking and storing run-requests, 2) synchronization tasks, 3) functions to protect hardware, 4) periodically executed functions, 5) various control functions, 6) helper functions for monitoring, and 7) functions for conversion from raw data to internal signals.

\subsection{Variant and structure analysis}
The two systems were managed as a single development track; however, during release, it was necessary to package the system for various applications, mechatronic hardware configurations, and brands. This process was internally called compiling and building the codebase into distinct products, and the need to package them into different configurations and brands contributes to higher maintainability effort and higher complexity. 

Scania's developers employed two primary strategies to address the resulting complexity: (1) splitting modules into variants and (2) incorporating variant handling directly within the modules' code. Splitting modules involved maintaining distinct .h and .c files with identical names but stored in separate directories within the file structure. While this approach reduced the complexity of individual files, it introduced the need to update multiple files during system maintenance. Table \ref{tab:variants} provides an overview of how these trade-offs were managed across the two systems.

\begin{table}[!ht]
    \centering
    \begin{tabular}{|c|c|c|}
    \hline
    \textbf{Variants} & \textbf{System A} & \textbf{System B} \\
    \hline
    1 & 5 & 41\\
    \hline
    2 & 13 & 67\\
    \hline
    3 & 12 & 6\\
    \hline
    4 & 5 & 4\\
    \hline
    5 & 8 & 2\\
    \hline
    6 & 7 & 1\\
    \hline
    7 & 3 & 0\\
    \hline
    8 & 0 & 0\\
    \hline
    9 & 5 & 0\\
    \hline
    \end{tabular}
    \caption{Module variants share the same boundary or interface, but the code is different, and the number of files in the module can vary over variants. The count in a system means the number of modules with that number of variants.}
    \label{tab:variants}
\end{table}

A comparison of the two systems shows significant differences in the number of module variants. Despite being smaller than System B in terms of lines of code and the total number of modules, System A has more module variants. Developer interviews quoted in the next section attribute this discrepancy to the broader range of hardware configurations controlled by System A. The diversity of mechatronic hardware versions and the associated complexity of control tasks render it impractical to encapsulate this complexity solely within the codebase. Consequently, the complexity is managed through the software build system. 

The organizational structures of the teams further highlight this distinction. Teams working on System A were organized around specific projects, whereas those managing System B assumed responsibility for modules independently of the project context. This organizational approach aligns with the more diverse control tasks required to manage System A's mechatronic hardware variations. Although System A's structure may initially seem to introduce additional complexity, it reflects a deliberate trade-off between code complexity and build system complexity. This approach balances these factors to effectively address the distinct challenges posed by each system.

\subsection{Thematic analysis}
A systematic analysis of the interview transcripts resulted in the development of a set of thematic codes documented in a codebook. The transcripts were then iteratively coded, during which new codes were identified, and overlapping codes were refined or merged. This iterative process led to continuous updates and refinements of the codebook. The finalized set of codes is presented in Table \ref{tab:codebook}.

\begin{table}[htbp]
    \centering
    \scriptsize
    \begin{tabular}{|p{2cm}|p{1.5cm}|p{5cm}|}
    \hline
    \textbf{Code} & \textbf{Code Shorthand} & \textbf{Description} \\
    \hline
    Definition of Maintainability & DEF\_MAINT & References to how interviewees define or conceptualize maintainability.\\
    \hline
    Complexity & COMPLX & Mentions of code complexity, complexity triggers, or specific statements about code structure that increases or decreases complexity.\\
    \hline
    Legacy Code & LEGACY & References to inherited or old code that is difficult to refactor due to risk or lack of documentation.\\
    \hline
    Testing & TEST & Anything about testing methods, unit tests, test cells, test coverage, challenges in testing, or the difference between testing "what it does" vs. "what it should do."\\
    \hline
    Requirements and Documentation & REQ\_DOC & Mentions of how---or whether---requirements are specified, documented, or communicated. Includes references to JIRA tickets or the absence of formal requirements.\\
    \hline
    Processes and Workflow & PROCESS & Mentions of the formal or informal steps needed to implement, approve, or release changes---e.g., CCB approvals, SOP tickets, risk analysis.\\
    \hline
    Hardware-Software Co-evolution & HW\_SW & Situations where the controlled hardware's changes drive software changes, or where the hardware is still in development while software must already be implemented.\\
    \hline
    Organizational Coordination & ORG\_COORD & References to interactions with other teams, specialized roles, or the challenges of distributing knowledge across multiple stakeholders.\\
    \hline
    Refactoring and Modularization & REFACTOR\_MOD & Mentions of modular design, code splitting, or attempts to isolate functionality to improve maintainability.\\
    \hline
    Regulatory / External Constraints & REGS & Mentions of R155, R156, or other regulatory requirements and how they influence design, logging, or maintainability.\\
    \hline
    Perceived Maintainability Over Time & PERCEPT\_TIME & Statements about whether the system’s maintainability has improved, declined, or remained the same over the past year.\\
    \hline
    Tools and Infrastructure & TOOLS\_INFRA & Mentions of version control, logging systems, or other tools that either help or hinder maintenance work.\\
    \hline
    Variant Handling / Product Diversity & VARIANT & Anything related to the explosion of variants, product lines, multiple configurations, or how branching logic affects maintainability.\\
    \hline
    Engineering trade-off & TRADEOFF & Mentioning active trade-off between different technical solutions to lower maintainability effort.\\
    \hline
    \end{tabular}
    \caption{Thematic codebook}
    \label{tab:codebook}
\end{table}

The application of thematic analysis to the coded transcripts resulted in the identification of several key themes. These themes, along with their associated codes, are as follows:

\begin{itemize}
    \item Higher number of variants lead to higher complexity and maintainability effort (VARIANT, COMPLX)
    \item Difficulty describing functional requirements impacts testing (TEST, REQ\_DOC)
    \item Risk awareness moderates legacy code refactorization (LEGACY, REFACTOR\_MOD)
    \item Parts of the perceived maintainability effort relate to organizational complexity (PERCEPT\_TIME, TOOLS\_INFRA, ORG\_COORD)
    \item Parts of the perceived maintainability effort relate to system external complexity (REGS, PROCESS, HW\_SW)
    \item Trade-off strategies for balancing maintainability effort (DEF\_MAINT, TRADEOFF)
\end{itemize}

The following sections detail the theme descriptions and present excerpts to show how the themes were constructed.

\subsubsection{Higher number of variants lead to higher complexity and maintainability effort}
Respondents highlighted that the number of supported software variants contributes to system complexity. Successfully managing these variants becomes part of the engineering challenge developers must solve, it is particularly evident when managing interactions between variants and external systems, as one developer explained:

\begin{quote}
   "When you have a lot of variants, and you have connection to outside of [the system] in any way, then it’s hard." 
\end{quote}

These developers need to solve the business problem of adding new variants while managing old ones. To effectively manage these business needs, developers need to make good design choices and adhere to processes to ensure risk-free product operation, which increases the maintenance effort. 

There was evidence of experimentation to try to encapsulate the complexity in the code to reduce the number of module variants, which indicates constantly balancing and optimizing the maintainability effort:

\begin{quote}
    "We started to try to reuse what was already written and just tweak it a bit and introduce variant handling in that code to handle the differences. But ... it ended up being too big differences."
\end{quote}

The developers' challenge was to successfully manage the added complexity from variant handling while simultaneously ensuring that safety and functional quality remained high. This was highlighted as the main part of the effort and required a deep understanding of the system.

\subsubsection{Difficulty describing functional requirements impacts testing}
The analysis of developer excerpts reveals that testing of functional requirements vehicle properties related to driver preference impacted the maintainability effort. Effective descriptions of these qualitative feeling-based functions were hard to formalize in written requirements. Instead, these types of requirements were conveyed via conversations and iterative explorations. While recognizing the effort needed to work with the qualitative requirements the developers also insisted on the value of documentation and requirements: 
\begin{quote}
    "If you don't have requirements, then it's difficult to know what to test against. How it’s meant to work."
\end{quote}

Mitigating risks related to understanding requirements included an organizational policy to pay for developers' driver's licenses, enabling them to spend a lot of time using the products. This gave them an intuitive feeling of the functional qualities that stakeholders conveyed. The developers highlighted this as a strength and something that helped them.

However, even though functional requirements describing the qualitative feel of vehicle behavior seem challenging, non-functional requirements were handled differently:
\begin{quote}
    "Risk mitigation requirements are different. That requirement handling is much more strict with test cases and follow-up."
\end{quote}

This implies that non-functional requirements were easier to convey because of their more quantitative nature. 

Testing functional requirements impacted the effort of maintaining the systems. Although conveying functional intent in written requirements was hard due to their qualitative nature, this problem was recognized and continuously improved. 

\subsubsection{Risk awareness moderates legacy code refactorization}
The analysis of developer excerpts highlights risk awareness when deciding whether to refactor legacy code. This type of code was described as contributing to complexity. When asked to define what such a function could be, one developer explained:
\begin{quote}
    "It’s implementing one huge function that does a lot of things. And it’s very difficult to get a good overview when you look at the code."
\end{quote}

Developers described how legacy code, built and modified over many years, could become deeply ingrained in systems, making it difficult to manage and evolve, leading to high maintainability effort. Because of this, the effort spent on refactoring was important. However, the risk of introducing errors during refactoring was highlighted, indicating that risk management was considered when deciding to refactor.

\begin{quote}
    "... it introduces some risk to refactor legacy code."
\end{quote}

On the other side of the balance were incentives for refactoring. For example, a function's intent could be lost, or coupling could be added to the system. If the developers didn't spend time refactoring, this could create confusion about the relationships between components. 

There were different reasons why legacy code had been structured as it had been. For example, the code could have been designed to accommodate practices and structures shaped by older project needs. 

While refactoring required effort, developers adopted strategies to mitigate risks and manage legacy complexity. One strategy was to split functionalities into separate modules managed by the build system, which allowed gradual improvements without disrupting stable components. Another strategy was using incremental steps towards a more modular structure and ensuring stable interfaces allowed parts of the system to evolve without widespread disruption:

\subsubsection{Parts of the perceived maintainability effort relate to organizational complexity}
The analysis of interview transcripts identifies several factors attributed to organizational complexity instead of software complexity; this influenced the perception of maintainability effort. 

Time pressure affected the perceived maintainability effort. Developers noted that the pace of project delivery needed to be considered and balanced with refinement or improvement of existing functionality:
\begin{quote}
    "But I mean, then it's hard to prioritize these kind of modularity and that work that enables future work, [like] maintainability, because the project pace is so high."
\end{quote}

The available tools and infrastructure also played a role in shaping maintainability effort perceptions. For example, developers described challenges with aligning development needs with changes to tooling infrastructure. Changing infrastructure and tooling required learning new systems and time spent refining ways of working, which added to the maintainability effort.

Coordination within and across teams impacted maintainability efforts, particularly in systems that support multiple products, customers, and variants. This added effort in terms of coordination overhead but was still highlighted as crucial for product delivery:
\begin{quote}
    "And we have the right contacts. Which is also really important... That you have contacts with the hardware people. ... it’s our delivery together with them."
\end{quote}

Interrelated factors such as time constraints, infrastructure limitations, and organizational coordination shaped the perceived maintainability effort. While developers reported these interactions as crucial for product development, they also conveyed that they added to the effort. 

\subsubsection{Parts of the perceived maintainability effort relate to system external complexity}
Compliance with external standards and regulations could introduce additional complexity since such regulations demand extensive testing, documentation, and certification.  Leaders found ways to lower this perceived effort by facilitating communication, for example, by arranging feedback meetings from tests.

The developers highlighted planning and synchronization as a source of perceived maintainability effort. One developer explained the need to provide good quality software to test prototypes of mechatronic hardware, which led to timing issues since hardware specifications could not be finalized until later in the development process:
\begin{quote}
    "The hardware is not frozen before the software needs to be frozen."
\end{quote}

To decrease the effort related planning and synchronization, and support effective collaboration between hardware and software teams, the organizational and architectural structures were modeled to suit the product:
\begin{quote}
    "You need to divide the software parts so that you have a collaboration with the hardware side."
\end{quote}

External factors such as regulatory compliance, processes, and evolving mechatronic hardware requirements impact perceived software maintainability efforts. While these factors were unavoidable, improved collaboration between teams, streamlining processes, and addressing hardware-software relations were identified as strategies that mitigated their effects.

\subsubsection{Trade-off strategies for balancing maintainability effort}
Developers often experienced the challenge of balancing maintainability with software evolution. The interviews revealed various strategies and considerations employed to achieve this balance.

Maintainability was widely understood as the ability to make changes or add new functionality with minimal risk to existing features. Developers emphasized clarity, modularity, and ease of testing as critical attributes:
\begin{quote}
    "Maintainability I would say it's the ease of adding functionality without risking functionality that's already in the files."
\end{quote}

Good maintainability also involved having a clear system architecture. Breaking code into smaller, manageable modules was also essential to improving maintainability. Developers highlighted the risks of overly large or complex functions:
\begin{quote}
    "Don't write too large functions or too large blocks of code. Instead, break it down into smaller pieces."
\end{quote}

Developers made clear that this trade-off regarding where to place the maintainability effort was not accidental but deliberate:
\begin{quote}
    "We’re replacing some [modules] in the build step… The trade-off is whether to create a variant or keep it common."
\end{quote}

Balancing maintainability required navigating trade-offs between modularity, variant handling, and reuse. Strategies like modularization, clear interfaces, and robust testing frameworks helped mitigate risks. Developers continuously evaluated which approach led to the lowest maintainability effort. By focusing on clarity and minimizing the cascading effects of changes, teams could manage these trade-offs effectively.

\subsection{Synthesis of quantitative and qualitative results}
The findings from this study highlight the multifaceted nature of maintainability in complex automotive software systems, demonstrating the interplay between quantitative code metrics and qualitative developer insights. This synthesis builds upon the theoretical framework and enriches it with new dimensions derived from the case study. Key results are summarized below:

\subsubsection*{Code Complexity and Its Impact on Maintainability}
Quantitative analysis of cyclomatic complexity, fan-in, and fan-out metrics across the studied systems revealed distinct patterns in complexity distribution. While most functions exhibited low complexity, legacy code and inherent product-specific challenges contributed to localized high complexity.

\subsubsection*{Complexity arising from team-external decisions}
A notable source of complexity and maintenance effort arose from the need to support diverse and complicated hardware control tasks. This complexity originated in organizational decisions to accommodate a wide range of controlled hardware configurations and requests from other brands within the corporate group. 

This need to support diverse mechatronic hardware configurations or move to new markets—adding regulatory requirements—drives internal and external complexity. Developers highlighted that this increased the maintainability effort. This highlights a recurring theme in the results: maintainability effort is not solely intrinsic to the code but also relates to business-driven decisions and the product domain.

\subsubsection*{Legacy Code and Risks}
Working with legacy code often meant working with complex functions. Although it was desirable to refactor, working with legacy code imposed risk and effort. Risk management and planning were considered crucial before attempting legacy code refactoring. Incremental modularization was highlighted as a preferable strategy over big-bang changes.

\subsubsection*{Trade-Offs and Developer Strategies}
Developers employed various strategies to balance complexity management and minimize maintainability efforts. Modularization, clear interface design, and selecting the best way to handle variants were key approaches. These trade-offs required conscious thought and deliberate design choices.

\subsubsection*{Perception of Maintainability}
Technical and organizational factors shaped developers' perceptions of maintainability. For example, time pressures and documentation challenges were cited as factors increasing maintainability efforts. Improving modularity and streamlining workflows were recognized as positive steps toward enhancing maintainability.

\subsubsection*{Proposed Extensions to the Theoretical Framework}
\begin{figure}[htbp]
    \centering
    \includegraphics[width=0.8\linewidth]{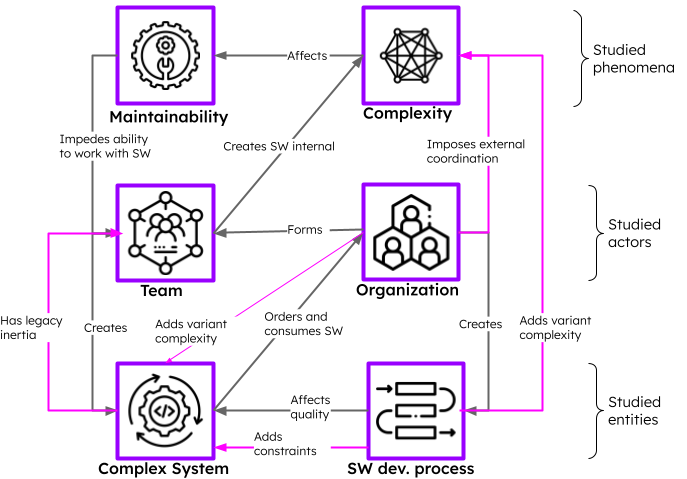}
    \caption{Expanded theoretical framework. Additions highlighted with pink lines.}
    \label{fig:FW_expanded}
\end{figure}

Building on the conceptual model presented in the theoretical framework, this study introduces the following additional elements:
\begin{itemize}
    \item \textbf{Variant Complexity:} A critical factor influencing maintainability, particularly in systems with diverse mechatronic hardware configurations.
    \item \textbf{Organizational Coordination:} The role of team structures and inter-team interactions in shaping maintainability outcomes.
    \item \textbf{Regulatory and Process Constraints:} External pressures necessitating trade-offs between maintainability effort and compliance.
    \item \textbf{Legacy Code Inertia:} The organizational and technical challenges associated with legacy systems and the strategies required to address them.
\end{itemize}
By integrating these elements, the theoretical framework provides a more comprehensive representation of the factors influencing software maintainability efforts in complex automotive systems. This expanded model offers a more realistic view of the theoretical framework applied in the context of this study. A visualization of the expanded theoretical framework is shown in Figure \ref{fig:FW_expanded}.

\section{Discussion}
\label{sec:discussion}
This study examined the relationship between sources of complexity and maintainability effort. The results are discussed in the following sections.

\subsection{RQ1: Developer perceived complexity}
In answer to RQ1, what aspect of complexity do developers highlight regarding maintainability effort? This study shows that developers face multiple sources of complexity and need to manage them: 1) the code itself can be complex due to complex hardware control tasks, 2) mechatronic hardware variant management in the code, or 3) there can be parts of the system that were built before the thinking-tools used for minimizing maintainability effort were introduced in the teams. 

There are also product management complexities that affect maintainability efforts. Hardware variant handling can be managed by creating module variants, creating product complexity that affects the maintainability effort. 

Furthermore, complexities also arise from system-external sources such as coordination complexity---e.g., to align plans---and communication complexity---e.g., understand qualitative personal-preference based functional requirements. External complexities, such as business decisions can also drive system internal complexity, e.g., decision to control a new variant hardware. 

Our findings regarding what developers perceive as complexities that affect maintainability efforts, align with existing literature, such as on the "technical debt" of legacy systems, where the accumulated cost of deferred maintenance undermines long-term system evolution \cite{Gordieiev}, and previous research emphasizing the influence of external organizational factors on software systems. For example, Bosch and Bosch-Sijtsema \cite{Bosch} discuss how market-driven decisions and product line strategies increase system variability and complexity. This result can explain why Mehwish et al.~\cite{Mehwish} found that code metrics predictive power for maintainability is still lacking. It fails to capture all the complexities that impact maintainability effort.

\subsection{RQ2: Code and variant complexity metrics}
Our analysis shows an alignment between measured complexity and the developers' answers in the interview study. The developers pointed out parts of their system that required more effort to maintain, and the measures confirm that these parts of the system are more complex as measured from a product complexity perspective. Our qualitative complexity measures revealed that the parts associated with complicated hardware control tasks, variant management, and legacy code were more complex, which aligned with developer perceptions.

Our findings align with other research regarding developer perceptions of complexity, such as in \cite{Buse, Holzmann, Mamun, Santos}. However, our work identified a stronger alignment between complexity metrics and maintainability effort. We also expand prior work by expanding the reasons developers provided for why the code was considered complex.

\subsection{RQ3: Team structure and interactions}
The results from the analysis showed that a well-thought-through team structure, e.g., aligning teams with high need of communication and alignment, such as embedded software teams and mechatronic hardware teams, reduces the maintainability effort. Other interactions required mitigations. One example was the communication of qualitative requirements. There, an organizational wide policy was implemented allowing driver's licenses for team members.

Our work expands the previous research \cite{Bosch, Concas, Lehman, Mubarak} by focusing on maintainability and identifying patterns that aid the joint development of complicated hardware control systems by software and hardware teams. The results show that aligning team and product structures for minimizing external complexity, by identifying the most complex communication paths, is a viable strategy for lowering maintainability effort.

\subsection{Emergent discovery of trade-off decisions}
During the analysis of the interview study, a theme emerged regarding trade-off decisions. The developers balanced design options and had clear intentions regarding how to minimize the maintainability effort. The main balancing point was increasing the module variant complexity or the code complexity. These findings suggest that code complexity metrics might need to be evaluated together with variant complexity measurements to understand the total complexity of a system.

These trade-offs resonate with prior studies on maintainability. Antinyan et al. \cite{Antinyan_maintainance_time} highlighted the difficulty of applying theoretical best practices in the face of real-world trade-offs, limiting the use of code complexity measures in industry. We have presented examples of trade-off decision-making, and traced strategies for iterative exploration of design choices that minimize maintainability efforts.

\subsection{Validity Considerations}
Several steps were taken to address validity concerns to ensure the credibility of the findings: 
\textbf{Construct Validity:} The use of both quantitative and qualitative data ensured a comprehensive understanding of maintainability. Metrics such as cyclomatic complexity, fan-in, and fan-out provided objective measures, while developer interviews offered insights into perceived maintainability challenges. This triangulation of methods enhanced the robustness of the findings. \textbf{Internal Validity:} Although causation is difficult to establish in case study research, the mixed-methods approach allowed for the exploration of relationships between code metrics, organizational practices, and perceived maintainability. For example, aligning quantitative metrics with qualitative reports of complexity in legacy code lends credibility to the identified relationships. \textbf{External Validity:} The study is limited to two embedded automotive software systems within a single organization, which constrains the generalizability of the results. However, including multiple products and systems within the same organizational context somewhat mitigates this limitation. Future research should explore maintainability in other domains to validate the transferability of these findings. \textbf{Reliability:} To ensure reliability, detailed documentation of the data collection and analysis process was maintained. Coding frameworks were iteratively refined, and cross-checking of codes was conducted to reduce bias in qualitative analysis. The quantitative analysis relied on established metrics and reproducible methodologies. \textbf{Researcher Bias:} While the authors sought to remain objective, the close collaboration with developers could introduce bias. The main author is also a long-time employee of the organization, and although this might be beneficial when interpreting the developers, preconceptions unknown to the author might still influence the results. Efforts to mitigate this included member-checking---sharing findings with participants for feedback---and using multiple data sources to validate interpretations. 

\section{Conclusions}
\label{sec:conclusions}
This paper shows that developers highlight different sources of system internal and external complexity as the main drivers of maintainability effort. Our findings suggest complexity measured with code and variability metrics align well with developer-perceived complexity. Furthermore, we found that team structures can mitigate certain types of external complexity in a system. We also found that developers make trade-off decisions in their designs to minimize maintainability efforts, often using experience-based heuristics as guiding thinking tools when making these decisions.

To conclude, the developer-experienced perceived maintainability effort is impacted by complexity, but only focusing on code complexity misses some sources of complexity. External complexities, like organizational and regulatory imposed ones, needs to be considered in order to fully understand maintainability effort. Our work shows what sources of complexities need to be considered when determining complexities that can be used as a proxy for the \textit{total} maintainability effort. To some extent explaining the lack of predictability power in current complexity code metrics, they seemingly do not capture the whole set of complexities that affect maintainability effort.

\subsection{Acknowledgments}
The authors would like to thank the developers, Scania CV AB, Software Center, and Vinnova.

\appendix
\section{Developer Interview Questions}
\label{sec:appendix_dev_questions}

The following questions were asked in the developer interviews:\newline

\textit{Developer experience questions:}
\begin{enumerate}
    \item Which parts of the system are you currently working on? For example, which files and or functions do you develop?
    \item How many years have you been working with this system?
    \item How many years of experience do you have as a developer?
\end{enumerate}

\textit{Maintainability questions:}
\begin{enumerate}
    \item When we talk about maintainability, what does this mean to you, i.e., how would you define it?
    \item If you were starting to work on a new system and wanted to know how hard the system was to maintain, what method would you use and what would you look for?
    \item With this in mind, how would you rate the system or part of the system that you are working with now in terms of maintainability?
    \item Any parts of the system or functions in particular that are extra tricky to understand, change, and maintain?
    \item How easy is it to test?
    \item If you need to make a change in the system, e.g., to add new functionality, can you describe the steps you need to take?
    \item How much effort would you say it takes to make such a change?
    \item If you need to fix a problem, could you describe how you perform analysis of the system?
    \item Would you consider this straight forward or requiring a lot of effort?
    \item How would you describe the modularity of the system?
    \item Can you reuse functionality in an easy manner, or do you often find yourself re-writing code that has already been written in other parts of the system?
    \item Are there any complexities outside of the code that makes the system hard to maintain? For example, when working with the system, do you often need to coordinate with other teams or individuals?
    \item Reflecting back over a year’s time, has the maintainability of the system become higher, lower, or stayed the same?
    \item Why do you think that is?
\end{enumerate}


\begin{thebibliography}{99}
\bibitem{Abílio} Abílio, R., Teles, P., Costa, H., \& Figueiredo, E. A systematic review of contemporary metrics for software maintainability. 2012 Sixth Brazilian Symposium on Software Components, Architectures and Reuse. IEEE, 2012.
\bibitem{Afshari} Afshari, M., \& Gandomani, T. J. (2021). A Typical Practical Team Structure and Setup in Agile Software Development. 2021 7th International Conference on Electrical, Electronics and Information Engineering (ICEEIE), Electrical, Electronics and Information Engineering (ICEEIE), 2021 7th International Conference On, 483–487. \url{https://doi.org/10.1109/ICEEIE52663.2021.9616743}
\bibitem{anda2007assessing} Anda, B. (2007). Assessing software system maintainability using structural measures and expert assessments,g in \emph{Proceedings of the 2007 IEEE International Conference on Software Maintenance}, pp. 204--213. \url{https://doi.org/10.1109/ICSM.2007.4362633}
\bibitem{Antinyan_hypnotized} Antinyan, V. (2021). Hypnotized by Lines of Code. Computer (Long Beach, Calif.), 54(1), 42–48. \url{https://doi.org/10.1109/MC.2019.2943844}
\bibitem{Antinyan_revealing} Antinyan, V. (2020). Revealing the complexity of automotive software. Proceedings of the 28th ACM Joint Meeting on European Software Engineering Conference and Symposium on the Foundations of Software Engineering, 1525–1528. \url{https://doi.org/10.1145/3368089.3417038}
\bibitem{Antinyan_maintainance_time} Antinyan, V., Staron, M., \& Sandberg, A. (2017). Evaluating code complexity triggers, use of complexity measures and the influence of code complexity on maintenance time. Empirical Software Engineering: An International Journal, 22(6), 3057–3087. \url{https://doi.org/10.1007/s10664-017-9508-2}
\bibitem{Antinyan_functions} Antinyan, V., Staron, M., Derehag, J., Runsten, M., Wikström, E., Meding, W., Henriksson, A., \& Hansson, J. (2015). Identifying Complex Functions: By Investigating Various Aspects of Code Complexity. Proceedings of 2015 Science and Information Conference (SAI), 879-. \url{https://doi.org/10.1109/SAI.2015.7237246}
\bibitem{Ardito} Ardito, L., Coppola, R., Barbato, L., \& Verga, D. "A Tool‐Based Perspective on Software Code Maintainability Metrics: A Systematic Literature Review." Scientific Programming 2020.1 (2020): 8840389
\bibitem{Bosch} Bosch, J., \& Bosch-Sijtsema, P. (2010). From integration to composition: On the impact of software product lines, global development and ecosystems. The Journal of Systems and Software, 83(1), 67–76. \url{https://doi.org/10.1016/j.jss.2009.06.051}
\bibitem{Buse} Buse, R. P. L., \& Weimer, W. R. (2010). Learning a Metric for Code Readability. IEEE Transactions on Software Engineering, 36(4), 546–558. \url{https://doi.org/10.1109/TSE.2009.70}
\bibitem{Coleman} Coleman, D., Ash, D., Lowther, B., \& Oman, P. (1994). Using metrics to evaluate software system maintainability. Computer (Long Beach, Calif.), 27(8), 44–49. \url{https://doi.org/10.1109/2.303623}
\bibitem{Concas} Concas, G., Marchesi, M., Murgia, A., Tonelli, R., \& Tempero, E. (2010). An Empirical Study of Social Networks Metrics in Object-Oriented Software. Advances in Software Engineering, 2010(2010), 1–21. \url{https://doi.org/10.1155/2010/729826}
\bibitem{Conejero} Conejero, J. M., Rodríguez-Echeverría, R., Hernández, J., Clemente, P. J., Ortiz-Caraballo, C., Jurado, E., \& Sánchez-Figueroa, F. (2018). Early evaluation of technical debt impact on maintainability. The Journal of Systems and Software, 142, 92–114. \url{https://doi.org/10.1016/j.jss.2018.04.035}
\bibitem{Elmidaoui} Elmidaoui, S., Cheikhi, L., Idri, A., \& Abran, A. Empirical studies on software product maintainability prediction: a systematic mapping and review. E-Informatica Software Engineering Journal 13.1 (2019): 141-202.
\bibitem{Gill} Gill, G. K., \& Kemerer, C. F. (1991). Cyclomatic complexity density and software maintenance productivity. IEEE Transactions on Software Engineering, 17(12), 1284–1288. \url{https://doi.org/10.1109/32.106988}
\bibitem{Gordieiev} Gordieiev, O., Gordieieva, D., Rainer, A., \& Pishchukhina, O. (2023, October). Relationship between factors influencing the software development process and software defects. In 2023 13th International Conference on Dependable Systems, Services and Technologies (DESSERT) (pp. 1-7). IEEE
\bibitem{Heitlager} Heitlager, I., Kuipers, T., \& Visser, J. (2007). A Practical Model for Measuring Maintainability. 6th International Conference on the Quality of Information and Communications Technology (QUATIC 2007), 30–39. \url{https://doi.org/10.1109/QUATIC.2007.8}
\bibitem{Henry} Henry, S., \& Kafura, D. (1981). Software Structure Metrics Based on Information Flow. IEEE Transactions on Software Engineering, SE-7(5), 510–518. \url{https://doi.org/10.1109/TSE.1981.231113}
\bibitem{Holzmann} Holzmann, G. J. (2016). Code Clarity. IEEE Software, 33(2), 22–25. \url{https://doi.org/10.1109/MS.2016.44}
\bibitem{Lehman} Lehman, M. M. (1980). On understanding laws, evolution, and conservation in the large-program life cycle. The Journal of Systems and Software, 1(3), 213–221. \url{https://doi.org/10.1016/0164-1212(79)90022-0}
\bibitem{ISO}ISO/IEC 25010:2023. 2023. Available online: \url{https://www.iso.org/obp/ui#iso:std:iso-iec:25010:ed-2:v1:en} (accessed on 28
May 2024)
\bibitem{Malhotra2} Malhotra, R., \& Lata, K. A systematic literature review on empirical studies towards prediction of software maintainability. Soft Computing 24.21 (2020): 16655-16677
\bibitem{Malhotra} Malhotra, R., \& Anuradha C. Software maintainability: Systematic literature review and current trends. International Journal of Software Engineering and Knowledge Engineering 26.08 (2016): 1221-1253.
\bibitem{Mamun} Mamun, M. A. A., Berger, C., \& Hansson, J. (2019). Effects of measurements on correlations of software code metrics. Empirical Software Engineering: An International Journal, 24(4), 2764–2818. \url{https://doi.org/10.1007/s10664-019-09714-9}
\bibitem{Mehwish} Mehwish, R., Mendes, E., \& Tempero, E.. A systematic review of software maintainability prediction and metrics. 2009 3rd international symposium on empirical software engineering and measurement. IEEE, 2009.
\bibitem{Mubarak} Mubarak, A., Counsell, S., \& Hierons, R. M. (2010). An evolutionary study of fan-in and fan-out metrics in OSS. 2010 Fourth International Conference on Research Challenges in Information Science (RCIS), 473–482. \url{https://doi.org/10.1109/RCIS.2010.5507329}
\bibitem{Runeson} Runeson, P., Host, M., Rainer, A., \& Regnell, B. (2012). Case Study Research in Software Engineering: Guidelines and Examples. In ELLIIT: the Linköping-Lund initiative on IT and mobile communication (1. Aufl.). Newark: Wiley. \url{https://doi.org/10.1002/9781118181034}
\bibitem{runeson2009guidelines} Runeson, P., \& Höst, M. (2009) Guidelines for conducting and reporting case study research in software engineering, \emph{Empirical Software Engineering}, vol.~14, no.~2, pp. 131--164.
\bibitem{Santos} Santos, G., Muzetti, I., \& Figueiredo, E. (2024). Two sides of the same coin: A study on developers’ perception of defects. Journal of Software: Evolution \& Process, 36(10), 1–18. \url{https://doi.org/10.1002/smr.2699}
\bibitem{staron2021automotive}
Miroslaw Staron, \textit{Automotive Software Architectures}. Springer, 2021.
\bibitem{weller2023developing}
Marcel Weller, Miles St{\"o}tzner, Floriment Klinaku, and Steffen Becker, 
\textit{Developing the Software of Future Cars: A Car DevOps Approach}, 
in \textit{Softwaretechnik-Trends Band 43, Heft 2}, Gesellschaft f{\"u}r Informatik e.V., 2023.

\end{thebibliography}
\end{document}